\documentclass[a4paper,fleqn,usenatbib]{mnras}
\usepackage{amssymb,amsmath,graphicx,psfrag,mathtools}
\usepackage[T1]{fontenc} 
\usepackage{ae,aecompl} 
\usepackage{url}
\usepackage{revsymb}
\DeclareFontFamily{U}{mathb}{\hyphenchar\font45}
\DeclareFontShape{U}{mathb}{m}{n}{
      <5> <6> <7> <8> <9> <10> gen * mathb
      <10.95> mathb10 <12> <14.4> <17.28> <20.74> <24.88> mathb12
      }{}
\DeclareSymbolFont{mathb}{U}{mathb}{m}{n}
\DeclareMathSymbol{\Earth}{3}{mathb}{"43}

\hypersetup{draft} 
\title[3 MHz Space Antenna]{3 MHz Space Antenna}
\author[J. I. Katz \& J. Krassner]{
	J. I. Katz$^{1}$\thanks{E-mail: katz@wuphys.wustl.edu} \& J. Krassner${^2}$ 
\\
$^{1}$Department of Physics and McDonnell Center for the Space Sciences,
Washington University, St. Louis, Mo. 63130 USA 
\\
$^{2}$Florham Park, N. J.
}
\date{Accepted XXX.  Received YYY; in original form ZZZ} 
\pubyear{2023} 
\date{\today}
\begin{document} 
\label{firstpage} 
\pagerange{\pageref{firstpage}--\pageref{lastpage}} 
\maketitle 
\begin{abstract}
	Little is known about the radio astronomical universe at frequencies
	below 10 MHz because such radiation does not penetrate the
	ionosphere.  A Cubesat-based {antenna} for the 1--10 MHz band
	could be rapidly and economically deployed in low Earth orbit.  
	When shielded by the Earth from Solar emission, it could observe weak
	extra-Solar System sources.  We suggest possible transient and
	steady sources, and application to study of the ionosphere itself.
\end{abstract}
\section{Introduction}
Low frequency ($\lesssim 25\,$MHz) {radio astronomy has been comparatively
little studied because it is difficult to observe these frequencies from the
ground.}  Science motivations for {its study} include very high redshift
cosmology, low frequency sky surveys, solar/space weather, possible
transient emissions from several classes of astronomical sources including
fast radio bursts (FRB), soft gamma repeaters (SGR), gamma-ray bursts (GRB)
and pulsars (PSR), and perhaps serendipitous discoveries.

The ionosphere reflects radiation at frequencies below a varying cutoff at
3--10 MHz at normal incidence, and at higher frequencies at grazing angles.
Scintillation {may be} prohibitively strong even at frequencies at which the
waves propagate.  These values depend on the Solar cycle and activity,
season and time of day, but largely preclude ground-based astronomy below 10
MHz.  The lowest frequency successful ground-based observations appear to
have been those of \citet{BP68} at 10.03 MHz, {of \citet{C76} at 10 MHz,} of
\citet{C79} at 5.2 MHz {and of \citet{E62} at 4.8 MHz}, although some
results at lower frequencies have been reported \citep{RE56,E57,E65,G69}.
{The abandonment of such low frequency ground-based observations more than
40 years ago reflects a consensus that they are not scientifically
promising.}

The ionosphere is not an obstacle to space-based observation.  In fact, its
existence is advantageous because it shields space-based instruments from
low frequency terrestrial electromagnetic interference while reflecting the
sky.  A few {antenn\ae\ for low frequency astronomical observations}
have been flown: Alouette \citep{H64} in LEO, {at a similar altitude to
that proposed here but in a high-inclination rather than equatorial orbit},
Radio Astronomy Explorer-1 (RAE-1) in a 5850 km orbit, {higher than the
LEO orbit contemplated here} \citep{A69,WAS71} and Radio Astronomy
Explorer-2 (RAE-2) in Lunar orbit \citep{A75}.  The prospects of space based
low frequency radio astronomy were reviewed in a conference \citep{KW90}.

{Several space instruments have observed Solar emissions at these low
frequencies, including WAVES on the deep space Wind and STEREO spacecraft
\citep{K05,WAVES,STEREO}.  The Sun Radio Interfermoeter Space Experiment
(SunRISE) \citep{SunRISE19,SunRISE22} is planned for geosynchronous orbit.
{Solar emission, including Type III bursts, might be a source of
confusion or interference for observations of dispersion-broadened
extra-Solar System bursts.  There can be no such confusion or interference
when an {antenna} is shielded from the Sun by the Earth, which occurs
for satellites in low Earth orbits (LEO) with a duty factor of about 40\%,
but never, or rarely, in deep space or geosynchronous orbits (GEO).}}

After a long period of somnolence, technical and scientific progress argue
for reviving space-based low frequency radio astronomy.  The technical
progress consists of the development of ``Cubesats'', satellites consisting
of one or more 10 cm cubes \citep{S18} {and of microelectronics capable
of sophisticated on-board data analysis within a small spatial envelope and
with minimal power}.  Cubesats are simple enough that they are built as
student projects and launch into low Earth orbit (LEO) may be free,
piggybacking on other launches.  The scientific progress consists of
the discovery \citep{L07} of FRB with durations ${\cal O} (\text{1 ms})$; 
the study of transients is a rapidly developing branch of radio astronomy.
The behavior of steady extra-Solar System radio sources at frequencies
$\lesssim 10\,$MHz is also unknown; strong steady sources might be located
by an {antenna} in LEO by occultation by the Earth's limb, or by the
Moon for favorably located sources \citep{A64}.

Observations at frequencies $\lesssim 10\,$MHz would constrain FRB radiation
mechanisms and environments (dense plasma prevents the escape of low
frequency radiation, {but at least one FRB is known to have a very clean
local environment \citep{F23,Z23}}).  Transients and rapidly varying sources
are advantageously observed from above the ionosphere at frequencies below
the ionospheric cutoff because in LEO the delay between direct and
ionospherically-reflected signals constrains their direction.  The peak of
the autocorrelation of the baseband signal at this delay determines a
source's zenith angle even without angular resolution.

{Uncharacterized but spatially smooth deviations of the ionosphere from
spherical symmetry would limit the accuracy of source localization by this
method, but would not greatly affect the strength of the reflected glint and
of the peak of the autocorrelation.  This peak would still distinguish a
transient or rapidly varying source from the background of steady sources as
well as from terrestrial interference, that may be significant even for low
frequency exo-ionospheric observation of steady sources \citep{A75}.
Measurement from an orbiting spacecraft of the zenith angles of two separate
transients from the same source would determine its location on the sky.}

\citet{R16} proposed a deep-space antenna array for low frequency radio
astronomy, {\citet{S16} proposed the CUbesat Radio Interferometry
Experiment (CURIE) to observe Solar radio bursts, \citet{Y22} described the
Low Frequency Interferometer and Spectrometer (LFIS) in Lunar orbit} and
\citet{B20} proposed the OLFAR (Orbiting Low Frequency Antennas for Radio
Astronomy) system involving hundreds or thousands of satellites, linked to
synthesize a large number of apertures.  \citet{L21} have placed a lander,
including a low frequency radio spectrometer, on the far side of the Moon
and the necessary data relay satellite in orbit.  These or similar projects
offer the prospect of great scientific return some time in the future, but
at high cost.

We propose a modest instrument that might, at less cost and sooner, perform
a preliminary survey of $\lesssim 10\,$MHz radio astronomy.  It would be
based on a Cubesat in LEO with two center-fed orthogonal half-wave dipole
antenn\ae; at a nominal frequency of 3 MHz ($\lambda = 100\,$m) these would
be extended to lengths $L = \lambda/4 = 25\,$m by centrifugal force in each
of four coplanar orthogonal directions.  This nominal frequency is suggested
because it is below the frequencies used by short-wave radio \citep{SW} and
likely also by over-the-horizon radar \citep{OTH}; {\it cf.\/} the signals
observed by RAE-2 in Lunar orbit \citep{A75}.

{The orbital altitude is chosen above the peak of ionospheric electron
density, where this density is low enough that it does not preclude
transmission of extra-terrestrial radiation at the frequency of observation.
A plasma frequency of 3 MHz (electron density $n_e = 1.15 \times
10^5\,$cm$^{-3}$) typically occurs at altitudes of 600--800 km
\citep{model}, depending on latitude, longitude, season, phase in the Solar
cycle and Solar activity.  It is desirable to be above this critical
altitude most of the time, so a nominal altitude $h = 1000\,$km is assumed.
Refraction by the ionospheric plasma is significant at that altitude, but
does not degrade the signal received by an {antenna} with little or no
angular resolution, such as the dipole antenn\ae\ considered.

At low frequencies the Sun is an intense source of background in a dipole
antenna.  The Earth shields this background whenever the {antenna} is in
its shadow.  At equinoxes it is shielded by the solid Earth a fraction
\begin{equation}
	\label{shield}
	f_{shield} = {1 \over \pi}
	\sin^{-1}{\left({R_\Earth \over R_\Earth+h}\right)} \approx 0.33
\end{equation}
of the time, where we have taken $h = 1000\,$km.

The ionosphere increases $f_{shield}$ because it increases the effective
(opaque) $R_\Earth$.  In addition, ionospheric refraction when the Sun is
near the {antenna's} horizon may further increase $f_{shield}$:
\begin{equation}
	f_{shield} = {1 \over 2} + {\phi - \psi \over \pi},
\end{equation}
where
\begin{equation}
	\phi = \cos^{-1}{\sqrt{1 - {\nu_{p\text{-}orb}^2 \over \nu^2}}}
\end{equation}
is the maximum angle of ionospheric plasma refraction, $\nu_{p-orb}$ is the
plasma frequency at the {antenna's} altitude and $\nu$ is the
{radiation} frequency.  The depression angle of the critical density
horizon
\begin{equation}
	\psi = \cos^{-1}{\left({R_\Earth + h_{crit} \over
	R_\Earth + h}\right)},
\end{equation}
where $h_{crit}$ is the altitude at which $\nu_p = \nu$; for observation to
be possible $h_{crit} < h$.

Both $\phi$ and $\psi$ are small angles for a satellite in LEO.  Averaging
over the year, $f_{shield}$ is multiplied by a factor $\approx
1-\varepsilon^2/4 \approx 0.96$, where $\varepsilon \approx 0.41\,$rad is
the obliquity of the Earth's equator.

In contrast, deep space or GEO observatories like FIRST, SURO-LC, DARIS,
Wind, STEREO, SunRISE and OLFAR are illuminated by the Sun; their purpose is
to observe it and its corona, but this background is a severe obstacle to
observations of weak extra-Solar System sources.  Refraction by
interplanetary plasma that broadens the arrival directions of low frequency
radiation by $\Delta\theta \sim 10^\prime \sim 3\,$mrad spreads its arrival
time by $\sim 1\,\text{AU} (\Delta\theta)^2/2c \sim 2.5\,$ms, limiting the
possible resolution of interferometry, even with long baselines.}
\section{The Antenna}
\begin{figure}
	\centering
	\includegraphics[width=\columnwidth]{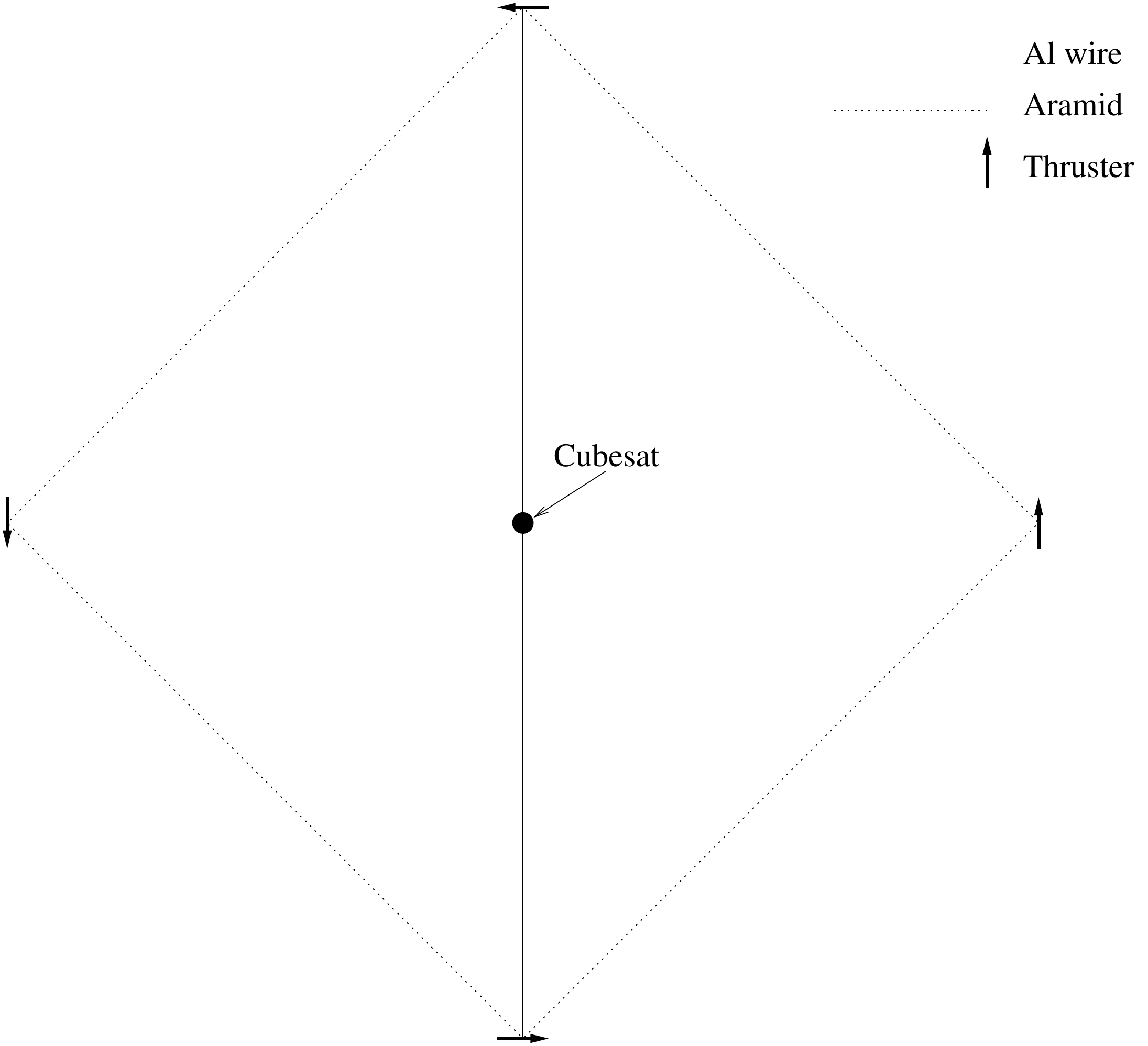}
	\caption{\label{telescope}Sketch of orbiting exoatmospheric
	radio telescope for observations at 1--10 MHz (numerical values
	in text apply for $\nu = 3\,$MHz).  Knudsen cell thrusters set
	the structure rotating, extending the wire antenn\ae.  The thrusters
	are aligned tangentially by ties to insulating aramid fibers,
	tightened by centrifugal force but that have no electromagnetic
	effects.}
\end{figure}

A minimal {system} is shown in Fig.~\ref{telescope}.  Two orthogonal
center-fed half-wave (at the nominal frequency of 3 MHz) dipole antenn\ae\
are extended from a Cubesat that contains amplifiers, data handling and
storage electronics and a {higher frequency} antenna for transmitting
data to a ground station.  Power is provided by Solar cells on the Cubesat.
Larger telescopes with {multiple} half-wave antenn\ae\ (separated by
distances $\sim \lambda/4$ to minimize capacitive coupling) could provide
some angular resolution.  {The configuration is similar to that of
Alouette \citep{H64}, but the construction and deployment made compatible
with a Cubesat.}

\subsection{Parameters}
A minimum antenna wire radius is set by the requirement that resistive
losses in the wire be small compared to its radiation resistance.  For a
resistance $\Omega$ in an aluminum wire of length $\lambda/2$ the wire
radius
\begin{equation}
	\label{r}
	r = \sqrt{\lambda/2 \over \pi \sigma_{Al}\Omega} \approx 0.21
	\sqrt{{\text{10 Ohms} \over \Omega}{\lambda \over \text{50 m}}}\
	\text{mm},
\end{equation}
where the conductivity of aluminum $\sigma_{Al} = 3.6 \times 10^7\,$mho/m
($\text{1 mho} = 1/\text{ohm}$).  This corresponds to 26 AWG (American Wire
Gauge).  The mass of the two half-wave antenn\ae\ is modest:
\begin{equation}
	\label{M}
	\begin{split}
		M &= 2\pi r^2 \rho_{Al} \lambda/2 \approx
		{\lambda^2 \rho_{Al} \over 2 \sigma_{Al} \Omega}\\
		&\approx 40 \left({\text{3 MHz} \over \nu} \right)^2
		\left({\text{10 Ohms} \over \Omega}\right)\ \text{g}.
	\end{split}
\end{equation}

{Past low frequency space antenn\ae\ have been partial cylinders or
tubes, extended, like a carpenter's rule, by mechanical stiffness.  This
requires a comparatively thick, stiff and massive antenna.  The advantage
of a centrifugally deployed wire antenna is that no mechanical thickness or
stiffness is required; it is extended by tension.  This permits a very thin
and low mass antenna, limited only the the requirement (Eq.~\ref{r}) that it
have sufficient electrical conductivity.}
\subsection{Sensitivity}
The detection threshold is a flux density
\begin{equation}
	\label{thresh}
	\begin{split}
		F_{thresh} &= {S \over N}{4 \pi \over \lambda^2}
		{k_B T_{rec} \over G_{rec}\sqrt{Bt_{int}}}\\
		&\approx 5 \times 10^4{S/N \over 10}
	{T_{rec}/3 \times 10^6\,\text{K} \over \sqrt{(Bt_{int}/10^6)}}\ 
	\text{Jy},
	\end{split}
\end{equation}
where $S/N$ is the required signal to noise ratio, $\lambda = 100\,$m the
radio wavelength, $T_{rec}$, the noise temperature of the receiver, {is
scaled to the Galactic synchrotron sky brightness at 3 MHz \citep{A69}},
$G_{rec}$ the telescope's antenna gain (taken as unity for a dipole), $B$
the receiving bandwidth and $t_{int}$ the integration time.

For a receiver bandwidth of {0.15} MHz, {5\% of the frequency}, and
an integration time of {3 s} (a pulsar with a {typical} duty factor
of {0.03} observed for 100 s); $Bt_{int} \approx 5 \times 10^5$.  The
sensitivity may be further improved by coherent processing at a
hypothetical or known pulsar period.  For a 1 ms FRB {$Bt_{int} = 1.5
\times 10^2$} and the detection threshold increases to {$\approx 3.5
\times 10^6\,$Jy}.

The flux densities of coherent radio sources increase rapidly with
decreasing frequency, and might be expected to be much greater at 3 MHz {
than in L-band}.  Pulsars typically have UHF and L-band spectral indices $\sim
-1.6$ \citep{L95}, so that extrapolation suggests flux densities $\sim 10^4$
times higher at 3 MHz than at 1 GHz, or $\sim 10^{10}\,$Jy for a source like
FRB 200428.  {Although the estimated detection threshold of $\approx 3.5
\times 10^6$\,Jy is about three times the {L-band} fluence of FRB
200428, it is much less that its extrapolated 1--10 MHz flux density}.
\subsection{Data Download}
A satellite in equatorial orbit passes over a near-equatorial ground station
once per orbit.  Data can be stored and downloaded with each passage over
the ground station.  The satellite would be within 2000 km for about 300 s
each orbit, implying a minimum data transmission rate of $\sim 10^9$ samples
per second for a receiver bandwidth of 3 MHz.

The required mean power to transmit a dual polarization base-band signal
sampled at a rate $2 \pi \nu_{obs}$ to a ground-based telescope of diameter
$D$ at a range $R$ is
\begin{equation}
	\begin{split}
		P_{tr} &= {64 \pi \nu_{obs} R^2 (S/N) k_B T_{data} \over
		D^2 G_{tr}}\\ &\approx {\text{60 mW} \over G_{tr}}
		{S/N \over 10} {T_{data} \over \text{30 K}}
		\left({R \over \text{2000 km}}\right)^2
		\left({\text{12 m} \over D}\right)^2,
	\end{split}
\end{equation}
where $S/N$ is the receiver signal-to-noise ratio, $T_{data}$ the data
receiver noise temperature and $G_{tr}$ is the transmitter gain.  {The
ALMA Band 1 receiver, operating at 35--50 GHz, somewhat above the Ka (26--40
GHz) band, has a specified noise temperature of 32 K, justifying scaling to
30 K, but even a system noise temperature of 100 K would imply a power
requirement of only $\sim 0.2\,$W.}

The required power is modest, even with a dipole transmitting antenna
($G_{tr} \approx 1$).  {Even a 1U Cubesat intercepts about 10\,W of
sunlight, and a larger satellite more.}  This could provide $>1\,$W of
photoelectric power with a duty factor ({\it q.v.\/} Eq.~\ref{shield})
$\approx 0.67$.

{The data transmission rate may be reduced by orders of magnitude if the
received signal is processed on-board, taking advantage of Moore's Law and
the revolutions in electronics since the era of Alouette, RAE-1 and RAE-2.
Rather than transmitting the base-band signal, it would only be necessary to
transmit the scientific information of interest.  This may be the total
received power as a function of time, the autocorrelation of the received
signal that indicates the presence of a transient event (because of the
delay between direct and ionospheric-reflected signal), or the location of a
steady source (again because the delay depends on its zenith angle).}
\section{Orbital Lifetime}
The mass of wire (Eq.~\ref{M}) is small compared to the typical mass
$\sim n\,$kg of a {nU} Cubesat, but the projected area of the antenn\ae,
each of length $\lambda/2$, is $2 \times 2 r \lambda/2 \approx 400\,$cm$^2$
(Eq.~\ref{r}), several times the projected area of a Cubesat.  Equating
the work done by atmospheric drag to the decrease in energy of the satellite
(noting that half the work done by gravity goes to increasing its kinetic
energy), the orbital altitude $h$ decreases at a rate
\begin{equation}
	\label{sinkrate}
	\begin{split}
		{dh \over dt} &= {4 C_d r L \rho_a \over M_{Cube}}
		\sqrt{GM_\Earth R_{orb}}\\ &\approx 0.022 {\text{1 kg} \over
		M_{Cube}} {\rho_a \over 10^{-16}\,\text{g/cm}^3}\
		\text{cm/s},
	\end{split}
\end{equation}
where $R_{orb}$ is the orbital radius (from the center of the Earth),
$M_{Cube}$ is the mass of the Cubesat, $\rho_a$ is the atmospheric density,
$M_\Earth \approx 6.0 \times 10^{27}\,$g is the mass of the Earth and the
drag coefficient $C_d$ is taken as unity.

The scale height of the atmosphere at altitudes of interest ({800--1200}
km) is about 20 km because of its elevated temperature.  As a result, the
characteristic orbital lifetime
\begin{equation}
	\label{orb}
	t_{orb} \equiv {\text{20 km} \over dh/dt} \approx 3\ {M_{Cube} \over
	\text{1 kg}} {10^{-16} \text{g/cm}^3 \over \rho_a}\ \text{y}.
\end{equation}
At these altitudes $\rho_a$ is sensitive to the Solar cycle and activity
\citep{J70,R71}, and also depends on time of day (but not much on season at
equatorial latitudes).  It is more useful to specify the air density than
the geometrical altitude, and it must be recognized that the orbital decay
time (Eq.~\ref{orb}) may decrease rapidly and unpredictably with Solar
activity.
\section{Deployment}
The antenn\ae\ must be extended by centrifugal force by setting the
telescope rotating.  Because of its small size, not much angular momentum
can be imparted to the Cubesat by forces applied to its surfaces, but even a
small initial angular momentum can begin the process of extension by
rotating the Cubesat.  As the antenn\ae\ extend, thrusters at their ends
produce increasing torques.

Several problems must be addressed:
\begin{enumerate}
	\item The thrusters must be simple, light, and cheap.
	\item The thrusters must continue to act over an extended time,
		perhaps hours or days, as the antenn\ae\ gradually extend,
		increasing their lever arms.
	\item The thrusters must be remain tangentially oriented.  The tiny
		torsional stiffness of the thin wire antenn\ae\ that
		connect them to the Cubesat is insufficient to align them.
		{Nor could tubular (or partial tubular) antenn\ae\ be
		both stiff enough and have walls thick enough for handling
		within the mass budget.}
\end{enumerate}

The first two problems are solved by using Knudsen cells \citep{G09} as the
thrusters.  A low vapor pressure compound, such as naphthalene, would
gradually escape through an aperture at one end of each cell, with its
recoil providing the thrust.

The third problem is solved by connecting the ends of the antenn\ae\ with 
fine electrically insulating fiber, such as aramid, as shown in
Fig.~\ref{telescope}, and fixing the Knudsen cells to the fibers.  Aramid
fibers are available as thin as 170 dtex (1 dtex is defined as a mass of 1
g/10 km) corresponding to a radius of about $60\,\mu = 0.006\,$cm (this is
also expressed as a length per unit mass $\text{Nm} = 60$, where 1 Nm is 1
m/g).  These fibers have the negligible total mass of about 2.5 g.  Each
fiber has a tensile strength of tens of N, orders of magnitude greater than
its tensile load at an angular rotation rate of 3.6/s (Eq.~\ref{omega}); it
is only necessary that the rotation rate be much greater than the orbital
angular frequency $\omega_{orb} \approx 10^{-3}\,$s$^{-1}$ in low Earth
orbit to maintain the geometry.

As the antenn\ae\ extend these fibers will also be made taut by centrifugal
force.  Thrusters tied to them would be aligned tangentially, so their
recoil forces spin up the entire system, keeping it taut and stable.
The moment of inertia of the four-armed (two $\lambda/2$ dipole antenn\ae)
telescope shown in Fig.~\ref{telescope} is
\begin{equation}
	I = {4 \pi \over 3} L^3 r^2 \rho_{Al} \approx 8 \times 10^7\
	\text{g-cm}^2,
\end{equation}
where Eq.~\ref{r} has been taken for the wire radius $r$.

Free molecular flow from Knudsen cells imparts an angular momentum
\begin{equation}
	{\cal L} = L m_p \sqrt{2 k_B T \over \pi m_g},
\end{equation}
where $m_p$ is the mass of propellant gas exhausted, $m_g$ its molecular
weight and $T$ its temperature.  $T$, and hence the vapor pressure and
evaporation time, are determined by the radiative properties of the outsides
of the Knudsen cells.  For naphthalene at 300 K the rotation rate
\begin{equation}
	\label{omega}
	\omega = {{\cal L} \over I} \approx 3.6 {m_p \over \text{10 g}}\
	s^{-1}.
\end{equation}
For $m_p = 10\,$g the load on the wire at the Cubesat is $\pi r^2 \omega
\rho_{Al} L^2/2 \approx 1.5\,$N and the tensile stress $\omega^2 \rho_{Al}
L^2/2 \approx 1.1 \times 10^8\,$dyne/cm$^2$, less than a tenth of the
tensile strength of aluminum.  The peripheral velocity $\omega L \approx 90
\,$m/s.

If the telescope plane is inclined at an angle $i$ to its orbital plane its
spin angular momentum and plane precess (as a result of the Earth's
gravitational torque) around its orbital angular momentum
at a rate
\begin{equation}
	\label{precess}
	\omega_{pre} = - {\omega_{orb}^2 \over \omega} \cos{i} \approx
	- 3.5 \times 10^{-7} \cos{i}\ \text{s}^{-1},
\end{equation}
or about one radian per month for the assumed parameters.  Spin precession
slews the broad dipole antenna pattern on the sky at an angular rate
$\omega_{pre} \sin{i}$ with angular amplitude $i$.  Significantly faster or
slower precession can be obtained by choice of $m_p$ and hence of the
rotation rate $\omega$.  The rotational plane of a telescope whose orbit is
not equatorial will also precess because of Earth's equatorial bulge, but
(if its spin and orbit are aligned) at a much slower rate than given by
Eq.~\ref{precess}.
\section{Sources}
There are no {known extra-Solar System} point sources of radiation in
the 1--10 MHz range, {but the mean emission of the Crab pulsar was
detected at 26.5 MHz with a flux density $\sim 1000\,$Jy by \citet{A64}
before its discovery as a pulsar!}

History has shown that observations in new regimes often discover new
phenomena.  For example, studies of atmospheric ionization discovered cosmic
rays, radio astronomy discovered active galactic nuclei (and inferred
supermassive black holes), time-resolved radio astronomy discovered pulsars,
X-ray astronomy discovered a zoo of neutron stars and stellar-mass black
holes, and pulsar astronomy led (in binary pulsars) to the confirmation of
the theory of gravitational radiation and to the archival discovery of FRB.
\subsection{Attenuation}
Galactic absorption is significant \citep{C79} at frequencies of a few MHz,
and at lower frequencies sets an effective horizon within the Galactic disc.
This need not preclude observation of old neutron stars that may radiate at
these lower frequencies, even though they are not detected in pulsar
searches at VHF and UHF frequencies.  There are $\sim 10^8$ neutron stars in
the Galactic disc, so that the nearest is likely at a distance of $\approx
20\,$pc.  The path from such a close source has an absorption optical depth
of only $\sim 0.2$ of that through the full thickness of the Galactic disc,
so the radiation from such a nearby source would be much less attenuated and
scattered than that from outside the disc.

The opacity at 3 MHz \citep{S62}
\begin{equation}
	\kappa_\text{3 MHz} = \begin{cases}
	1.05 \times 10^{-17} n_e^2\ \text{cm}^{-1} & \text{T} = 100\
		\text{K} \\
		1.8\phantom{0} \times 10^{-21} n_e^2\ \text{cm}^{-1} &
		\text{T} = 10^4\ \text{K},
	\end{cases}
\end{equation}
where the electron density $n_e$ (in cm$^{-3}$) is assumed to come from
singly ionized species.  In a weakly ionized ($n_e = 0.03\,$cm$^{-3}$) cool
(100 K) cloud the absorption length is $\sim 30\,$pc and varies nearly
$\propto T^{3/2}$, while in a warm ($10^4$ K) ionized intercloud medium
($n_e = 0.01\,$cm$^{-3}$ in pressure equilibrium with a cool neutral cloud
with $n_\text{H} = 1\,$cm$^{-3}$) the absorption length is $\sim 150\,$kpc.

Cool clouds may be opaque, but cosmic ray electrons within them produce an
internal source of radio radiation.  Condensation into clouds likely 
increases their cosmic ray density and magnetic field, so they may be net
emitters in comparison to the extra-Galactic background.  The warm ionized
intercloud medium is transparent, transmitting the extra-Galactic
background.
\subsection{Dispersion and Broadening}
Dispersive time delays are large at low frequencies (where the dispersion
measure DM has been scaled to convenient values for FRB):
\begin{equation}
	\label{Deltat}
	\Delta t = 2.3 \times 10^5 \left({\text{DM} \over
	\text{500 pc-cm}^3} \right) \left({\text{3 MHz} \over \nu}\right)^2\
	\text{s},
\end{equation}
and
\begin{equation}
	{d \Delta t \over d\nu} = - 1.54 \times 10^5 \left({\text{DM} \over
	\text{500 pc-cm}}\right) \left({\text{3 MHz} \over \nu}\right)^3\
	{\text{s} \over \text{MHz}}.
\end{equation}
As a result, a signal that is impulsive at its source is greatly broadened
at 3 MHz after propagating through a dispersive medium characteristic of FRB
or even of Galactic pulsars, {and even at the plausible $\text{DM} \sim
0.3\text{--}1\,$pc-cm$^{-3}$ of the closest neutron stars.}

{In addition to dispersion, multipath scattering broadens transients
\citep{KMNJM}.  This can be a large effect at low frequency because it
scales approximately as the $-4$ power of frequency.  However, the
scattering measure (its coefficient) varies by several orders of magnitude
among sources, and may be very small for nearby {objects, such as the
closest old ``dead''} pulsars.  Although the number of pulsars detected at
higher frequencies is smaller, old neutron stars, dead at higher
frequencies, might pulse in the 1--10 MHz band.}

Dispersion {and multipath scattering between the source and the Solar
System} do not affect the geometrical time delay $\delta t$ between the
direct and the ionospherically reflected signals (the ionospheric dispersion
measure from the altitude at which 3 MHz radiation is reflected to deep
space is only $\sim 10^{-6}\text{--}10^{-7} \,$pc-cm$^{-3}$).  Hence the
autocorrelation of the signal received by an {antenna} in LEO would show
a peak at the geometrical time delay $\delta t$ (Eq.~\ref{theta}).

{This peak is broadened by the dispersive time delay (Eq.~\ref{Deltat}).
As a result, the autocorrelation of a signal of width $\tau$ at its source
is broadened by ${\cal O}(\tau/\Delta t)$, and the signal to noise ratio of
detection of the autocorrelation peak is reduced by ${\cal O}(\sqrt{\tau/
\Delta t}) \sim 10^{-4}$ for the dispersion measure of a cosmological or
even distant Galactic plane source (the observed Galactic FRB 200428 had
$\text{DM} = 332.7\,$pc-cm$^{-3}$; \citep{Bo20}).  However, FRB 200428 had
an observed L-band fluence $1.5\,$MJy-ms, about a million times more intense
than a typical cosmological FRB detected with $S/N \ge 10$.  Extrapolation
to $< 10\,$MHz is speculative, but most astronomical radio sources (unless
self-absorbed, which a coherent source would not be) have negative spectral
slopes (higher flux density at lower frequency), making detection of the
autocorrelation peak at least plausible.}

As the {antenna} moves in its orbit, a source's zenith angle $\theta$
and autocorrelation peak $\delta t$ vary.  As discussed in
Sec.~\ref{reflection}, with knowledge of the orbit the time dependence of
an autocorrelation peak may be measured, indicating the presence of a
variable (dispersed burst) source as well as its location on the sky; only
one orbiting {antenna} would be required.  Additional positional
information may be obtained from the times of occultation by the Earth's
limb.
\subsection{Transient and Variable Sources}
{Interstellar and intergalactic dispersion and scattering make the
detection of transients difficult at low frequency.  However, the
autocorrelation of the baseband voltage peaks at a lag corresponding to the
delay between the direct and ionospheric-reflected paths and may enable the
detection of even heavily broadened and dispersed transients.  This is only
possible for an {antenna} in LEO, for which the direct and reflected
signals have similar strength.  {Pulsars are known with $\text{DM} <
3$ pc-cm$^{-3}$ \citep{PW89,ATNF}, implying little dispersion and scatter
broadening.  More numerous and even} closer old and slow pulsars, ``dead''
at higher frequencies, {likely have even} smaller dispersion and scatter
broadening.  They might be detectable in the 1--10 MHz band, as might novel
interstellar plasma processes.}

It is not known if pulsars emit radiation in the 1--10 MHz band, {but
pulsars with periods from 0.25 s to 1.27 s have been detected at frequencies
as low as 25 MHz \citep{PW89}, demonstrating their emission at low
frequencies and suggesting that detectable emission may extend to even lower
frequencies.}  The number of old PSR that might be observable at frequencies
below 10 MHz may far exceed the number detected in past searches.  Detection
would constrain PSR radiation mechanisms because radiation frequencies are
determined by the energy of the radiating particles, the local magnetic
field, and the plasma processes by which they emit.  Detection would also
provide new information about PSR population statistics and spindown history.

Coherent emission from FRB has been detected at frequencies as low as 110
MHz \citep{P21} and 120 MHz \citep{PM21}, with no evidence of a low
frequency turnover or cutoff.  \citet{CN71} speculated about coherent
supernova emission at low frequencies and \citet{UK00} about coherent radio
emission by gamma-ray bursts.
%
\subsection{Steady Galactic Sources}
Electron cosmic rays emit incoherent but nonthermal radio synchrotron
radiation, and provide a well-understood background.  The absorption of this
radiation by interstellar ionized gas diagnoses the spatial and temperature
distribution of that gas \citep{EH66,WAS71,A75,C79}.  The proposed system
would extend these observations to lower frequencies where the interstellar
plasma absorption, varying as the $-2$ power of frequency, is greater.  It
may also be possible to infer the absorption along paths to very bright
discrete sources, such as Cygnus A, a diagnostic of the interstellar medium
along its line of sight, by observing the change of its contribution to the
sky-integrated signal as it enters or leaves Earth occultation. 
\section{Ionospheric Reflection}
\label{reflection}
The ionosphere is a good reflector at 3 MHz, so the antenn\ae\ would observe
the reflection of a transient by the ionosphere as well as the direct
signal.  {A flat-ionosphere approximation is justified because for the
suggested orbit the antenna is a height $h \sim 300\,$km above the
reflecting layer.  That layer is approximately spherical with a radius
$R \approx 7000\,$km, about 600 km above the Earth's surface.  If the
electron density contours are smooth and horizontal, the difference
$\delta \ell$ between the direct and reflected signal paths is determined by
the source's zenith angle $\theta$ and elevation $\Delta = \pi/2 - \theta$:
\begin{equation}
	\label{theta}
	\delta \ell \approx h \left({1-\cos{2\Delta} \over
	\sin{\Delta}}\right) = 2 h \sin{\Delta}.
\end{equation}

If the signal is a brief pulse, a dipole antenna, with broad angular
acceptance, would observe two pulses separated by a time interval $\delta t
\approx \delta \ell/c$, localizing the source to a circular arc on the sky.
Simultaneous detection by two telescopes would confine the source location
to the two intersections of two arcs.  If the source fluctuates then the
autocorrelation of its base-band signal may have a peak at the lag $\delta
t$ even if the emission extents over a time $> \delta t$.}

Use of Eq.~\ref{theta} requires knowing the instantaneous height of the
reflective layer, which varies with Solar activity.  It can be
measured in real time if the antenna emits a pulse and receives its
reflection.  Rapidly varying or impulsive sources within the Solar System,
anthropogenic or natural, may be localized by this method.  {In fact, the
reflective surface of the topside ionosphere may not be accurately flat
because of ionospheric turbulence, but may vary in an uncertain manner,
limiting the accuracy of this method of source localization.} 
\section{The Topside Ionosphere}
\label{topside}
The spatial structure of the topside ionosphere \citep{BSR76,P20,P22} may
be probed by observing the reflection of 3 MHz radiation from a {strong
steady natural source or an artificial beacon.  Natural sources include
the radio galaxies Cen A and Cyg A and the SNR Cas A.  If the reflective
layer is tilted by gravity waves (or otherwise), the source's effective
elevation varies, and can be inferred from the phase difference and
interference between the reflected and direct signals from steady sources
of known direction.

The low Galactic latitude Cas A ($b = -1.96^\circ$) may be observable only
at frequencies $\gtrapprox 10\,$MHz because of interstellar absorption
\citep{S23}, that may also be significant for Cyg A ($b = 5.6^\circ$).  At
larger zenith angles (lower elevations) the ionosphere is reflective at
higher frequencies, less absorbed by interstellar plasma, and grazing
reflection at these frequencies can also be used to study the topside
ionosphere.

In the simplest possible model the ionosphere is static (except for
the effect of the diurnal variation of the Solar ionizing radiation) and its
density contours are spherical (horizontal in the flat-ionosphere
approximation).  However, the elevation $\Delta$ (and zenith angle $\theta$
vary as the satellite moves in its orbit, so the path difference $\delta
\ell$ and phase difference $2 \pi \delta \ell/\lambda$ vary with time,
comparatively rapidly.  For a steady source the received power will
oscillate as the phase difference between these two paths changes as
$\Delta$ varies with the satellite motion.  {This is the same principle
as that of the Sea Interferometer \citep{BS53} of early radio astronomy,
itself a long-wave realization of Lloyd's mirror.}

For an ideal static ionosphere the frequency of oscillation is determined
by the satellite's orbit and the direction to the source.  Ionospheric
oscillations (tilts of the reflecting surface, which is the critical density
surface for sources at the zenith but is higher for sources at nonzero
zenith angles) will manifest themselves as deviations of these oscillations
from the predictions of the flat-ionosphere model.  This would require only
measurement of the received power, averaged over a fraction of the
oscillation period (typically seconds), rather than processing or
transmission of base-band data.

The elevation $\Delta$ of a source at a known Right Ascension $\alpha$
and Declination $\delta$ as viewed from a point in an equatorial orbit
($\delta_{orb} = 0$) with geocentric Right Ascension $\alpha_{orb} =
\Omega t$, where $\Omega$ is the orbital angular velocity, is given by
\begin{equation}
	\cos{\Delta} = \cos{\delta}\cos{(\alpha-\Omega t)}.
\end{equation}
The amplitude of the sum of the direct and reflected fields, and hence of
the received power, oscillates with a period $P_{osc}$.  The rate of change
of the delay between the two signals is
\begin{equation}
	\begin{split}
		d\delta t &= {2h \over c}d\delta\sin{\Delta}
		= {2h \over c}\cos{\Delta}d\Delta\\
		&= {2h \Omega\over c}\cos^2{\delta}
	\cos{(\alpha-\Omega t)}\sin{(\alpha-\Omega t)}dt,
	\end{split}
\end{equation}
or
\begin{equation}
	{d \delta t \over dt} = {h \Omega \over c} \cos^2{\delta}
	\sin{2(\alpha-\Omega t)}.
\end{equation}

A full cycle of this oscillation occurs after a time
\begin{equation}
	d\delta t = {2\pi \over \omega}
\end{equation}
where $\omega$ is the angular frequency of the radio radiation.  This
condition is met after an interval
\begin{equation}
	\label{posc}
	P_{osc} = {2\pi/\omega \over d\delta t/dt} = {cP_{orb} \over
	\omega h}{1 \over \cos^2{\delta}\sin{2(\alpha-\Omega t)}},
\end{equation}
where $P_{orb}$ is the orbital period.  For $P_{orb} = 100\,$min, $\omega =
2 \times 10^7$\,s$^{-1}$ (3 MHz) and $h = 300$\,km, the numerical
factor is about 0.3 s.  As many as ${\cal O}(10^4)$ cycles of this
oscillation could be observed in the approximately half-orbit (3000 s)
during which an astronomical source can be observed, so that ionospheric
tilts ${\cal O}(10^{-4})$ radian might be detectable.

The approximately 5\% bandwidth of a half-wave dipole is wide enough that
the oscillation phase varies by many radians across the band, so that 
the band would have to be divided into channels to be analyzed separately.
The oscillation frequency (or $P_{osc}$) of the received power is described
by few tens of bits of data because the oscillation has a frequency of a few
Hz.  Transmitting these stored data, even for hundreds of frequency
channels, would not require much bandwidth.}

The delay also requires correction for propagation through layers in which
the electron density is high enough that the signal group velocity is
significantly less than $c$, and its path is bent by refraction.  Because
the electron gyrofrequency is 1--2 MHz, the effect of the geomagnetic field
is significant and the ordinary and extraordinary modes must be considered
separately within the ionosphere.

{This analysis depends on the assumption that the ionospheric density
contours are smooth, undisturbed by turbulence.  The failure of these
predictions (variation of the received intensity that is not periodic with
the period of Eq.~\ref{posc}) would therefore be a novel probe of topside
ionospheric turbulence that is not addressed by existing ionospheric
diagnostics \citep{Sc23}.}
\section{Discussion}
Even without angular resolution, it would be possible to probe the Universe
in this unexplored frequency band:
\begin{itemize}
	\item A single dipole antenna (or co-located orthogonal dipoles)
		would be sensitive to transients.  Time intervals between
		direct and ionospherically reflected signals would provide
		positional information.
	\item A dipole antenna would measure a sky average (weighted by its
		gain) temperature, yielding information about the
		interstellar medium not obtainable in any other manner.
		Models of the interstellar medium predict the antenna
		temperature of a dipole, and can be tested by its
		measurement.
	\item A dipole antenna observing the reflection of an exospheric
		beacon would measure the variability of the topside
		ionosphere.
\end{itemize}

Angular resolution would provide additional information:
\begin{itemize}
	\item A single dipole telescope (or two orthogonal dipoles) whose
		rotation axis is not parallel to Earth's would precess, even
		in equatorial orbit.  This would sweep its dipole beam
		pattern across the sky, providing some angular resolution of
		steady emission like that of the interstellar medium.
	\item Resolution could be obtained by aperture synthesis with a
		larger telescope comprising multiple dipoles.  This could
		resolve interstellar cloud structure.
	\item Aperture synthesis using multiple, widely spaced, dipoles in
		equatorial orbit could narrowly constrain the locations of
		transients, but would be poorly matched to the broad angular
		scales of interstellar clouds.  More closely spaced dipoles
		would be a better match to that angular structure.
		Knowledge of the dipoles' locations (although not
		necessarily active station-keeping) to allow for
		unpredictable differences in atmospheric drag would be
		required; this could be provided by GPS.
\end{itemize}

An equatorial orbit would permit data downloads to a single ground station
every orbit, {minimizing data storage and download rate requirements.
The orbit of such a satellite, locked to the Earth's equatorial bulge, would
not precess.  If rotating in the Earth's equatorial plane, its spin would
also not precess.}  

Large, low frequency space-based antenna\ae\ are plausible candidates for
demonstration of in-space manufacturing and assembly. Such structures would
need to be quite large ($\sim\,$km) in order to provide even degree-scale
resolution at these long wavelengths.  Before such a costly and technically
challenging demonstration it would be prudent to develop a ``pathfinder''
Cubesat-scale telescope to explore the signal characteristics to be observed
by a larger instrument. 

{An extended antenna wire poses a collision risk for other satellites.
Its effective cross-section for a 1 m satellite would be $\sim 50\,$m$^2$.
This is much larger than the satellite's $\sim 1\,$m$^2$ cross-section for a
small piece of space debris, but the proposal is only for one {antenna}
while lethal debris are numerous.  If the wire is} thin enough collision
might not be catastrophic to the satellite.  Once the mission is over, the
antenn\ae\ could be detached from the Cubesat.  No longer centrifugally
extended, a slight (unavoidable) intrinsic curvature would crumple the thin
wire into a compact tangle, with much reduced cross-section.

The mutual collison risk of multiple 3 MHz {antenn\ae}, such as might be
deployed to make an interferometer, would be minimized if all were in
equatorial orbits, with rotation axes parallel to the Earth's and each
other's.  Then their mutual collision cross-section would be proportional
only to the first power of their antenna length, rather than quadratic.

\section*{Acknowledgments}
We thank Joseph Lazio, David Palmer {and anonymous referees} for
calling our attention to earlier work on this subject.
\section*{Data Availability}
This theoretical work generated no original data.

\label{lastpage} 
\end{document}